\documentclass[aps,prl,twocolumn,superscriptaddress]{revtex4} 

\usepackage{amsfonts}
\usepackage{amsmath}
\usepackage{amssymb}
\usepackage{graphicx}%
\usepackage{bm}

\usepackage{mathrsfs}
\begin{document}


\newcommand{\Co}{Ba(Fe$_{1-x}Co_x$)$_2$As$_2$}
\newcommand{\ie}{{\it i.e.}}
\newcommand{\eg}{{\it e.g.}}
\newcommand{\etal}{{\it et al.}}


\title{Anisotropy of the Optimally--Doped Iron Pnictide Superconductor Ba(Fe$_{0.926}$Co$_{0.074}$)$_2$As$_2$}


\author{M.~A.~Tanatar}
\email{tanatar@ameslab.gov}
\affiliation{Ames Laboratory, Ames, Iowa 50011, USA}

\author{N.~Ni}

\affiliation{Ames Laboratory, Ames, Iowa 50011, USA}
\affiliation{Department of Physics and Astronomy, Iowa State University, Ames, Iowa 50011, USA }

\author{C.~Martin}

\affiliation{Ames Laboratory, Ames, Iowa 50011, USA}

\author{R.~T.~Gordon}

\affiliation{Ames Laboratory, Ames, Iowa 50011, USA}
\affiliation{Department of Physics and Astronomy, Iowa State University, Ames, Iowa 50011, USA }

\author{H.~Kim}
\affiliation{Ames Laboratory, Ames, Iowa 50011, USA}
\affiliation{Department of Physics and Astronomy, Iowa State University, Ames, Iowa 50011, USA }

\author{V.~G.~Kogan}
\affiliation{Ames Laboratory, Ames, Iowa 50011, USA}

\author{G.~D.~Samolyuk}
\affiliation{Ames Laboratory, Ames, Iowa 50011, USA}

\author{S.~L.~Bud'ko}
\affiliation{Ames Laboratory, Ames, Iowa 50011, USA}
\affiliation{Department of Physics and Astronomy, Iowa State University, Ames, Iowa 50011, USA }

\author{P.~C.~Canfield}
\affiliation{Ames Laboratory, Ames, Iowa 50011, USA}
\affiliation{Department of Physics and Astronomy, Iowa State University, Ames, Iowa 50011, USA }

\author{R.~Prozorov}
\email{prozorov@ameslab.gov}
\affiliation{Ames Laboratory, Ames, Iowa 50011, USA}
\affiliation{Department of Physics and Astronomy, Iowa State University, Ames, Iowa 50011, USA }

\date{29 December 2008}


\begin{abstract}

Anisotropies of electrical resistivity, upper critical field, London penetration depth and critical currents have been measured in single crystals of the optimally doped iron pnictide superconductor Ba(Fe$_{1-x}$Co$_x$)$_2$As$_2$, $x$=0.074 and $T_c \sim$23~K. The normal state resistivity anisotropy was obtained by employing both the Montgomery technique and direct measurements on samples cut along principal crystallographic directions. The ratio $\gamma_{\rho} = \rho_c /\rho_a$ is about  4$\pm$1 just above $T_c$ and becomes half of that at room temperature. The anisotropy of the upper critical field, $\gamma_{H} = H_{c2,ab} /H_{c2,c} $, as determined from specific heat measurements close to $T_c$, is in the range of 2.1 to 2.6, depending on the criterion used. A comparable low anisotropy of the London penetration depth, $\gamma_{\lambda}=\lambda_{c}/\lambda_{ab}$, was recorded from TDR measurements and found to persist deep into the superconducting state. An anisotropy of comparable magnitude was also found in the critical currents, $\gamma_j=j_{c,ab}/j_{c,c}$, as determined from both direct transport measurements ($\sim$1.5) and from the analysis of the magnetization data ($\sim$3). Overall, our results show that iron pnictide superconductors manifest anisotropies consistent with essentially three-dimensional intermetallic compound and bear little resemblance to cuprates. 

\end{abstract}

\pacs{74.70.Dd,72.15.-v,74.25.-q}





\maketitle



\section{Introduction}

Discovery of the iron-arsenide family of high critical temperature superconductors \cite{Hosono} naturally raises the question about the relation between them and the cuprates \cite{Science_Cho}. These materials share many common features. Both materials are layered with electronically active Cu-O and Fe-As layers alternating with buffer layers of different chemical composition. The $3d$ electronic orbitals of copper and iron make the main contribution to the electronic bands close to the Fermi energy. As a result of this layered structure, the cuprates reveal highly anisotropic electronic properties, with the ratio of conductivities along and perpendicular to the conducting layer, $\gamma_{\rho} \equiv \rho_c /\rho_a$, varying from about 50 in YBa$_2$Cu$_3$O$_7$ \cite{YBCO_anisotropy} to above 10$^3$ in Bi$_2$Sr$_2$CaCu$_2$O$_{10}$ \cite{BiSCO_anisotropy} at optimal doping. This high anisotropy is a reflection of a two-dimensional Fermi surface, as found experimentally in semiclassical and magnetoquantum oscillations \cite{HusseyTl2201,Louis2}. High anisotropies are also found in the lower \cite{Hc1cuprates} and upper critical fields \cite{Hc2cuprates} and in the critical current density \cite{Jcranisotropycuprates}. 

For the iron arsenides, early band structure calculations have also suggested a two-dimensional electronic structure \cite{iron_BandS}. High anisotropy of the resistivity, with $\gamma_{\rho} \sim $100, was reported for the non-superconducting parent compounds BaFe$_2$As$_2$ \cite{BaFeresaniz} and SrFe$_2$As$_2$ \cite{SrFeresaniz}, as well as for Co-doped BaFe$_2$As$_2$ \cite{Coaniz}. Contrary to this, a relatively low anisotropy is found in the upper critical field of all studied iron arsenide compounds close to optimal doping \cite{anisotropy,NiNi,WangHc2anisotropy,Martin,Singleton,Mielke,Welp1,Welp2,Yamamoto08,anisotropy_japanese,NiNiCo}. The angular dependence of resistivity as a function of magnetic field also suggests a rather small anisotropy in NdFeAs(O,F) \cite{Wenaniz}. The anisotropy evolves with doping, showing a two-fold change between the underdoped ($x<$0.074) and overdoped ($x>$0.074) regions \cite{NiNiCo}. This suggests that the anisotropy may be an important parameter to characterize superconductivity in the iron pnictides.

The anisotropy of the upper critical field, $\gamma_H \equiv \frac{H_{c2,ab}}{H_{c2,c}}$, and of the London penetration depth, $\gamma_{\lambda} \equiv \lambda_c / \lambda _{ab}$, are linked in the region of validity of the Ginzburg-Landau (GL) theory of phase transitions close to $T_c$, also for isotropic gap and in the dirty limit, $\gamma_H \sim \sqrt{\gamma_{\rho}}$ \cite{linkedgammas,Gorkov,Pokrovsky,KRap}. Gross violation of these relations and very high values of $H_{c2}$ \cite{highHc2} can indicate, for example, paramagnetic \cite{Clogston} or some exotic \cite{Japanese_Deguchi} mechanism of superconductivity suppression already in the very vicinity of $T_c$. This situation is realized in two-dimensional organics for magnetic fields applied parallel to the superconducting planes \cite{Kovalev}, in the spin-triplet superconductor Sr$_2$RuO$_4$ \cite{Deguchi} and in the heavy fermion CeCoIn$_5$ \cite{Hc2CeCoIn5}. In all of these cases the temperature interval of the validity of GL theory is very small and the temperature dependence of $\gamma_H$ at any sizable field does not follow GL predictions. 

In this work we have undertaken comprehensive characterization of the anisotropy of Ba(Fe$_{1-x}$Co$_x$)$_2$As$_2$, $x$=0.074, with $T_c \approx$23~K. Our choice was motivated by the availability of high quality single crystals and very good reproducibility of their properties between different groups for all doping levels \cite{Athena,NiNiCo,Ian}. We have found that the anisotropies of the upper critical field, electrical resistivity, London penetration depth and critical current, which were determined using different measurements, agree with each other and show values much lower than in the cuprates, indicating that the salient physics associated with superconductivity in these two families of compounds may be different.

\section{Experimental}

Single crystals of Ba(Fe$_{1-x}$Co$_x$)$_2$As$_2$ were grown from FeAs/CoAs flux from a starting load of metallic Ba, FeAs and CoAs mixed in the proportions 1:3.6:0.4, as described in detail elsewhere \cite{NiNiCo}. Crystals were thick platelets with sizes as big as 12$\times$8$\times$1 mm$^3$ and large faces corresponding to the tetragonal (001) plane. The cobalt content in the crystals was determined by wavelength dispersive X-ray electron probe microanalysis to be $x$=0.074. 
The crystal quality of the samples was confirmed with X-ray Laue measurements on single crystals, which found resolution limited narrow peaks, see \cite{NiNiCo} and \cite{RTGordon} for details. 

It was shown in previous study that correct determination of the sample resistivity is not a simple problem for the iron arsenides \cite{NiNiCo}. Due to softness of the materials, their cutting and shaping into transport samples inevitably introduces cracks, which affect the effective geometric factors of the sample. This represents an especially serious problem for measurements with current along the $c$ axis. A strong tendency to exfoliate prevents the cutting of samples with $c \gg a$. As we will show later, partial cleaving by exfoliation is one of the most likely reasons for the unusually high anisotropy, as found in previous studies \cite{BaFeresaniz,SrFeresaniz,Coaniz}. 

Samples for electrical resistivity and critical current measurements with current flow along the [100] $a$-axis in the tetragonal plane ($\rho _a$ and $J_{c,a}$) were cut into bars of 5*0.12*0.025 mm$^3$ ($a\times b\times c$), as described in Ref.~\cite{vortex}. Samples for the Montgomery technique measurements (see below) of the resistivity anisotropy ratio, $\gamma_{\rho}$, as well as  for electrical resistivity and critical current measurements with current flow along the tetragonal $c$ axis ($\rho_c$ and $J_{c,c}$ ) were cut into (0.2-0.5)*(0.2-0.5)*(0.1-0.5)mm$^3$ ($a\times b\times c$) bars.

Contacts to the samples were made by attaching silver wires with a silver alloy, resulting in an ultra low contact resistance (less than 100 $\mu \Omega$). Measurements of $\rho_a$ and $J_{c,a}$ were made in both standard 4-probe and 2-probe configurations and gave identical results, see \cite{vortex} for details. Measurements of $\rho_c$ and $J_{c,c}$  were made in the two-probe sample configuration.  Even for the nominal 2-probe measurements, a 4-probe scheme was used to measure the resistance down to the contact to the sample, i.e. the sum of the actual sample resistance $R_s$ and contact resistance $R_c$ was measured. Since $R_s \gg R_c$, this represents a minor correction. This can be directly seen at temperatures $T<T_c$, where $R_s =$0 and measured resistance represents $R_c$ (see Fig.~\ref{rhoa} and Fig.~\ref{rhoc} below).

\begin{figure}
	
		\includegraphics[width=8.5cm]{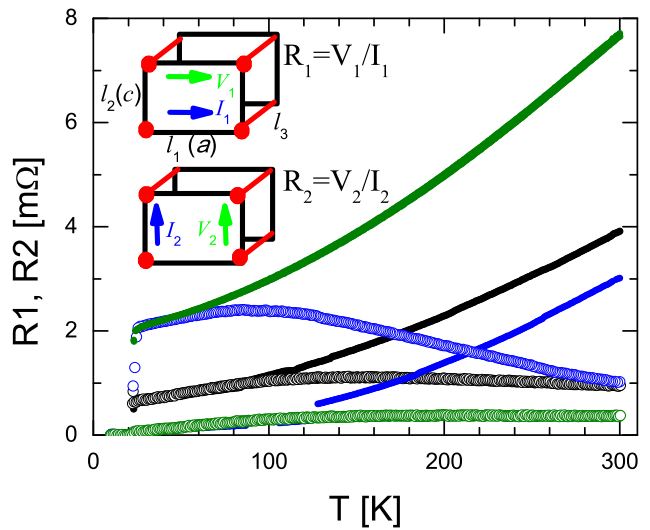}
	\caption{\label{Montgomery_raw} Inset: Arrangement of electrical contacts in the 4-probe Montgomery technique for determination of electrical resistivity anisotropy. The sample is cut into a rectangular prism, with the sides of its bases $l_1$ and $l_2$ oriented along the principal directions $1$ and $2$ of the conductivity tensor, $a$ and $c$ crystallographic directions in our case. Contacts (red lines) are put over the whole length of the sample in the third direction, $l_3$ ($\parallel$a). Two successive 4-probe measurements are made with current (blue arrows) flowing between contacts on one side of the sample and voltage (green arrows) measured on the other side of the sample, to determine resistance values $R_1$ (current along $l_1$) and $R_2$ (current along $l_2$).
Main panel: Temperature dependence of resistances $R_1$ (small solid symbols) and $R_2$ (large open symbols) for three different samples of Ba(Fe$_{1-x}$Co$_x$)$_2$As$_2$, $x$=0.074. 
}
\end{figure}

In 1961 Wasscher \cite{Wasscher}, based on van der Pauw calculations \cite{vanderPauw_scaling}, found that the current distribution in a sample with dimensions of $l_1$ and $l_2$ along principal directions of the resistivity tensor, $\rho_1$ and $\rho_2$, is equivalent to that of an isotropic sample with dimensions $l_1 $ and $\l_2 \sqrt{\rho _2/\rho_1}$ \cite{vanderPauw_scaling}. This scaling transformation is used in the Montgomery technique to map measurements on samples of a known geometry and an unknown anisotropy onto those in isotropic samples, where this ratio can be calculated \cite{Montgomery1,Montgomery2}. Two successive 4-probe resistance measurements are made using the contact configuration shown in the inset in Fig.~\ref{Montgomery_raw}. First, current flows along the $l_1$ direction, in our case corresponding to the $a$-axis in the plane, producing a voltage drop on the opposite side of the sample to determine the resistance $R_1=V_1/I_1$. Second, the direction of the current is rotated by 90 degrees along the $l_2$ direction (along the tetragonal $c$-axis in our experiment), and the resistance R$_2$ is determined. 
The ratio of the measured resistances, $R_1/R_2$ is exponentially sensitive to the ratios of sample dimensions, $l_1/l_2$, and is used for determination of the resistivity anisotropy \cite{Montgomery1}. 
Since the whole idea of the Montgomery technique is based on a homogeneous current distribution in the sample, the structural integrity of the sample plays a crucial role for measurements of this type. 


Analysis of the measured anisotropy signal was performed assuming the precise position of the contacts at the corners of the sample and neglecting their size in the basal $ac$ plane of the rectangular prism. The contacts were extending along the sides perpendicular to the basal plane, so a thin slab approximation was used in the analysis. Since the actual size of the contacts is not negligible as compared to the sample circumference in the $ac$ plane, and their positions can deviate from ideal, this brings sizable errors into the anisotropy measurement. By reproducibility of the results on three measured samples, we estimate the error to be on the order of $\pm$ 50\% for the anisotropy ratio. 

Transport critical current measurements were performed by measuring $I ? V$ characteristics at fixed temperatures, see \cite{vortex} for details. 
Samples for $J_{c,a}$ measurements were mounted with GE-varnish on an insulating oxide substrate to serve as a heat sink and contacted with either multiple silver wires or silver foil to create large area contacts providing an additional heat sink. Samples for $J_{c,c}$ measurements were sandwiched between two silver foils and thermally grounded to large silver heat sinks. Currents up to 2 A were generated in a commercial PPMS measuring system from {\it Quantum Design}.
Magnetic critical current densities were estimated by using the Bean model \cite{Bean_model} from the magnetization measurements performed in a {\it Quantum Design} SQUID magnetometer and independently from the profiles of the magnetic induction, measured by utilizing a magneto-optical method \cite{MO}.

Measurements of the penetration depth were performed by using a tunnel diode resonator (TDR) technique \cite{TDR_technique}. This technique is capable of resolving changes in the penetration depth of about 1 \AA. Details of the measurement technique are described elsewhere \cite{Prozorov2006}. In brief, a properly biased tunnel diode compensates for losses in a tank circuit, so it is self-resonating at a frequency of $f =1/(2 \pi \sqrt{LC})\sim 14$ MHz.  A sample is inserted into the coil on a sapphire rod. The change of effective inductance causes a change in the resonant frequency. This frequency shift is proportional to the dynamic magnetic susceptibility of the sample, $\chi $. Knowing the geometrical calibration factors of the circuit, we obtain $\lambda \left( T,H\right) $ as described in Ref. \cite{Prozorov2006}. A major advantage of this technique is a very small AC excitation field amplitude ($\sim 20$ mOe), much lower than $H_{c1} \sim$ 100 Oe.  This means that the TDR technique only probes, but does not disturb the superconducting state. Other advantages are high stability and excellent temperature resolution ($\sim 1$ mK).

The upper critical field $H_{c2}$ was determined from the onset of the superconducting transition in the TDR measurements as well as from measurements of the specific heat, taken in a {\it Quantum Design} PPMS. The same sample was used in the specific heat, DC magnetization measurements of the critical currents and in the magneto-optical imaging. In the configuration with $H\parallel ab$, the sample was aligned parallel with the field using a precisely cut sapphire substrate. Two samples were studied and they have shown similar properties.

Band structure calculations have been done within the full potential linearized augmented plane wave (FLAPW) approach \cite{FLAPW} within the local density approximation (LDA) \cite{LDA}. The mesh of 31x31x31 $\vec k$-points was used for the Brillouin zone integration and Fermi surface plot. We have used experimental lattice constants for the BaFe$_2$As$_2$ \cite{latticeconstants} and Ba(Fe$_{0.926}$Co$_{0.074}$)$_2$As$_2$ \cite{NiNiCo}. The Fermi velocities were calculated using the Bolz-Trap \cite{BolzTrap} package.

\section{Results}

\subsection{Anisotropic resistivity measurements}

\begin{figure}
	
		\includegraphics[width=8.5cm]{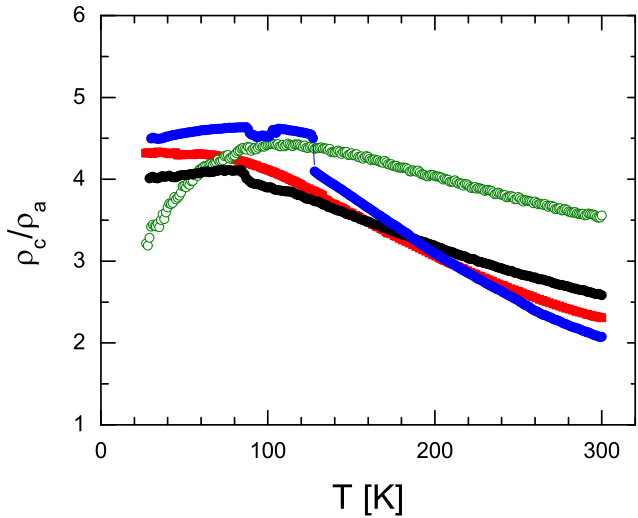}
	\caption{\label{Montgomery_anisotropy} Temperature dependence of the ratio of resistivities $\rho_c / \rho_a$ as determined from the Montgomery technique. For comparison we show the temperature dependence of the ratio as determined from independent $\rho_a$ and $\rho_c$ measurements (red line).
}
\end{figure}

In Fig.~\ref{Montgomery_raw} we show the temperature dependence of the experimentally measured resistances $R_1$ (current along the plane, solid symbols) and $R_2$ (current perpendicular to the plane, open symbols) for 3 different samples. At room temperature, $R_1$ is higher than $R_2$ in all the samples, as expected for samples of larger in the plane dimension and small anisotropy. The resistivity anisotropy, $\gamma_{\rho}= \rho_c /\rho_a$, was deduced from the Montgomery procedure and is shown in Fig.~\ref{Montgomery_anisotropy}. There is an overall general agreement in the temperature dependence of both resistances for the three samples measured, however, with variation of $R_{1}/R_{2}$ due to its exponential dependence on the ratio of the sample dimensions. The anisotropy $\gamma _{\rho}$ varies between 2.2 and 3.5 at room temperature and increases on cooling  approximately by a factor of two, reaching (3 to 5) at $T_c$. Resistance jumps seen in two samples most likely indicate that the samples undergo partial cracking on cooling. Since the Montgomery technique heavily relies on the idealized current distribution in the sample, it is impossible to ascertain that the ratio is determined correctly at temperatures below which the crack formation happens. However, at room temperature and at temperatures down to the appearance of cracks, the data seems to be quite reliable and well reproducible. Of the three samples, the most reliable measurement was done on sample \#1 (black curves), where $R_1$ and $R_2$ are comparable over the entire temperature range, implying good compliance with requirements of Montgomery analysis. 
Cracks affect both $R_1$ and $R_2$, suggesting current redistribution in the sample, although resistance values do not change dramatically.


\subsection{ Artifact of high anisotropy and different temperature dependence}

\begin{figure}
	
		\includegraphics[width=8.5cm]{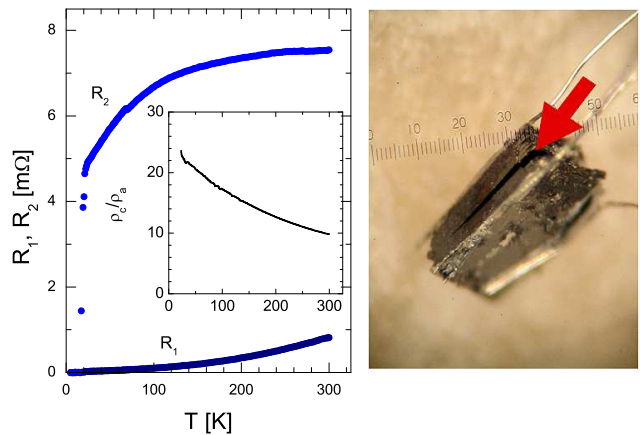}
	\caption{Left panel: Temperature dependence of the raw resistance data $R_1$ and $R_2$ (main panel) and of the resistivity anisotropy ratio $\rho_c / \rho_a$ (inset) for sample 4. Note very different temperature dependence of $\rho_a / \rho_c$ as compared to samples 1-3, in Fig.~\ref{Montgomery_anisotropy}, despite a very similar temperature dependence of individual values of $R_1$ and $R_2$. Right panel: photograph of the sample after two month storage. On degradation, sample revealed clear crack perpendicular to the $c$-axis, which was not noticeable after sample preparation. 
}
	\label{Montgomery_artifact}
\end{figure}

In Fig.~\ref{Montgomery_artifact} we show the $R_1$ and $R_2$ data from yet another sample, measured in the Montgomery configuration. At room temperature this sample shows $R_2 \sim 7*R_1$, despite comparable sample dimensions, suggestive of notably higher anisotropy, as can be seen in the inset of Fig.~\ref{Montgomery_artifact}. Interestingly enough, $R_1$ and $R_2$ individually reveal a similar temperature dependence to those found in the other 3 samples, however, the ratio of the measured resistivities, as determined from the analysis, gives a very different temperature dependence as compared to samples 1-3. As we will show later, the anisotropy ratio determined from direct resistivity measurements shows the same temperature dependence as the ratio determined from the Montgomery technique measurements on samples 1-3. Based on this qualitative difference we conclude that the data for this sample is not representative. 

We should point, however, that even with such notably different measured $R_1$ and $R_2$, the anisotropy ratio increases by a factor of 5 or so, but notably distorts the temperature dependence. This stresses the need for direct resistivity measurements with the current along principal directions of the conductivity tensor.

To get insight into the possible reason for this behavior, we inspected the sample with unusual anisotropy ratio. The right panel of Fig.~\ref{Montgomery_artifact} shows the sample after two months storage. Due to sample degradation, the cracks, which were not noticeable originally, developed and became noticeable. This particular sample almost split into two pieces.


\subsection{Temperature dependence of in-plane and inter-plane resistivity} 

Since Montgomery resistivity measurements allow for the correct determination of the ratio of resistivities, but not their individual temperature dependences, we need to perform direct measurements of at least one of the components. To check the consistency of the obtained results, we have independently measured the resistivity for both current flow directions. 

\begin{figure}
	
		\includegraphics[width=8.5cm]{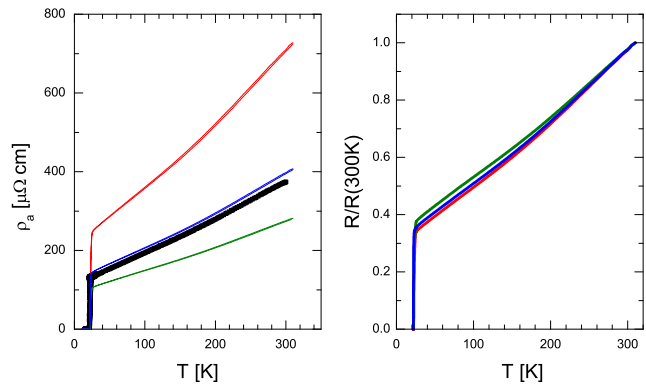}
	\caption{Temperature dependence of $\rho_a $ measured on 4 different samples from 3 different batches. Left panel- actual measurements, right panel- the data normalized to the room temperature values. }
	\label{rhoa}
\end{figure}

\begin{figure}
	
		\includegraphics[width=8.5cm]{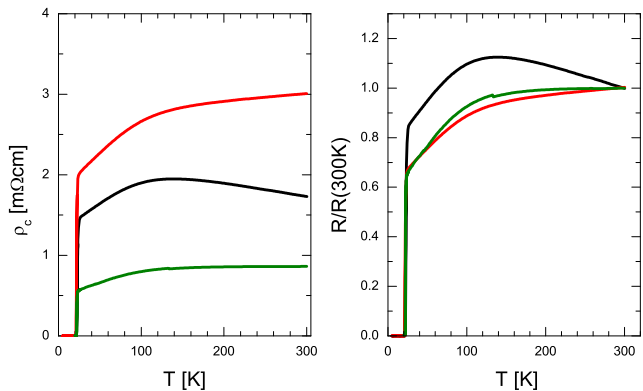}
	\caption{Temperature dependence of $\rho_c $ measured on 3 samples. Left panel- actual measurements, right panel- the data normalized to the room temperature values. }
	\label{rhoc}
\end{figure}

The data for $\rho_a(T)$ are shown in Fig.~\ref{rhoa}. The data are taken on samples from 3 different batches. Note that the scatter of resistivity values at room temperature is far above the error bar in the geometric factor determination. On the contrary, as shown in the right panel, the shape of the temperature dependence remains the same, with $\rho_a(T)$ varying roughly linear in $T$ over  the entire temperature range between room temperature and $T_c$. The data can be actually matched even better allowing for slight variation of the residual resistivity $\rho_0$. This is consistent with the results of our study using different sample cutting/contact making technique \cite{NiNiCo}. 

This similarity of the shape of the temperature dependence together with a large difference in the inferred resistivity values clearly shows that the problem comes from incorrect determination of the geometric factor. Two types of defects can be invoked to explain this discrepancy. Inclusions of FeAs flux usually happen between layered crystallites. Cracks tend to exfoliate layers. In both cases the effect will be more visible for $c$-axis transport, but will affect the $\rho_a$ measurements as well. However, since FeAs is a metal, it is hard to imagine that its own contribution to the conductivity will not affect the shape of the resistance temperature dependence. No evidence for FeAs flux inclusions is found also in x-ray studies \cite{NiNiCo}. 

The temperature dependence of $\rho_c$ taken on three different samples is shown in Fig.~\ref{rhoc}. Due to small sample size along $c$-axis we were only able to perform two probe resistivity measurement. As discussed in Methods section above, the contact resistance is negligibly small. 
The three curves are similar, which is better seen when the data are normalized by room temperature values, Fig.~\ref{rhoc}, right panel. They show an extended range of weak variation in $\rho_c(T)$ down to approximately 100~K followed by a roughly linear decrease below. A factor of 4 difference in the $\rho_c$ value implies variation of effective sample cross-section due to cracks. 

\begin{figure}
	
		\includegraphics[width=8.5cm]{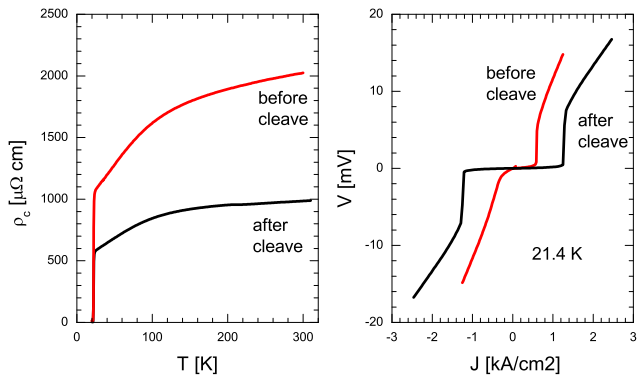}
	\caption{Left panel Temperature dependence of $\rho_c $ measured on the same sample after initial preparation and after cleaving the part of the sample and remaking contacts. Right panel shows V-J curves for the same sample at 21.4~K, revealing increased critical current density after sample cleaving. }
	\label{cleave}
\end{figure}

To test the possible effect of cracks on $\rho_c(T)$ measurements we performed two successive measurements of the resistivity and of the critical current on the same sample. After an initial run with $\rho_c(T)$ and $J_c$ measurement, we perceived that $\rho _c$ was potentially artificially high. To test this the sample was cleaved into two pieces by pulling contact wires and applying small force parallel to the plane. A new contact was made on fresh cleaved surface, and measurements were repeated. The resultant $\rho_c(T)$ and $V-J$ curves are compared in Fig.~\ref{cleave}. As is clear from the figure, cleaving decreases resistivity of the sample (beyond changing only geometric factor) and increases critical current density. This is consistent with the sample initially having a crack perpendicular to the $c$-axis, giving rise to the high $\rho _c$ value. In addition, this clearly shows that samples with the lowest measured resistivity $\rho_c$ should be used for evaluation of resistivity anisotropy. 

For $\rho_a$, the effect of cracks due to exfoliation is more complicated. Here cracks not only affect measured resistivity but can disrupt connection between current and voltage contacts. As a result this can either increase (if current flows across the crack) or decrease (if voltage contacts are disconnected from current path) measured resistance values \cite{TTF-TCNQ}. Because of this, we have excluded from consideration the extreme curves. As an additional criterion for the selection, we have used the data for the sample with the highest critical current density, which would obviously exclude samples with cracks. 

In Fig.~\ref{Montgomery_anisotropy} we show a comparison of the temperature dependence of the anisotropy ratio from independent measurements of $\rho _a (T)$ and $\rho_c (T)$, and the ratio as measured by using the Montgomery technique. Taking into account that these data are taken on different samples in very different measurement conditions and a sizable amount of uncertainty in geometric factors, the agreement between the two independent anisotropy determinations is remarkable. 

To evaluate error bar for the determined anisotropy ratio, we need to estimate a spread in resistivity values for $\rho_a$ and $\rho_c$. As we have shown above, simple averaging of $\rho_c$ is not meaningful. As extreme case scenarios we take minimum (maximum) resistivity $\rho_a$ and $\rho_c$ and come to the anisotropy $\rho_c/\rho_a$ between 1.2 and 10 at room temperature. This covers all the range of Montgomery technique determinations, and is notably lower than the number reported in previous study \cite{Coaniz}.

\subsection{ Anisotropy of the upper critical field}. 

In order to compare the anisotropy of the electrical resistivity to the anisotropy of $H_{c2}$, we need to determine $H_{c2}(T)$ near $T_c$. Two previous studies performed resistive determination of $H_{c2}$ \cite{Yamamoto08,NiNiCo} on samples of Ba(Fe$_{1-x}$Co$_{x}$)$_2$As$_2$ with similar $T_c$ and composition. In Ref.~\cite{Athena} $x$=0.1 according to the starting load, similar to our samples, which actually corresponds to 0.074 \cite{NiNiCo}. Both found low $\gamma_H$. Close to $T_c$ $\gamma_H$ is about 2 \cite{Yamamoto08} or 2.5 to 3.2 \cite{NiNiCo}. On cooling much below $T_c$ it decreases to 1.5 at $\sim T_c /2$ \cite{Yamamoto08} or 1.5 to 2 at 0.7$T_c$ \cite{NiNiCo}. The slopes, $dH_{c2}/dT$, were found to be 4.5 T/K and 2.4 T/K for $H \parallel ab$ and $H \parallel c$, respectively \cite{Athena}, however, these are strongly criterion dependent \cite{NiNiCo}. This projects to a resistivity anisotropy $\gamma _{\rho} ~\sim \gamma_H^2$ of 4 to 9, which is in reasonable agreement with our findings. However, resistive determination of $H_{c2}$ is strongly criterion-dependent and potentially subject to contributions of superconducting fluctuations and current percolation. This can notably complicate determination of $H_{c2}$ at the lowest fields, where changes are small. With this in mind, we have decided to use specific heat measurements as a bulk probe of the $H_{c2}$, revealing simultaneously sharp and easy to recognize feature of the superconducting transition. 

\begin{figure}
	
		\includegraphics[width=8.5cm]{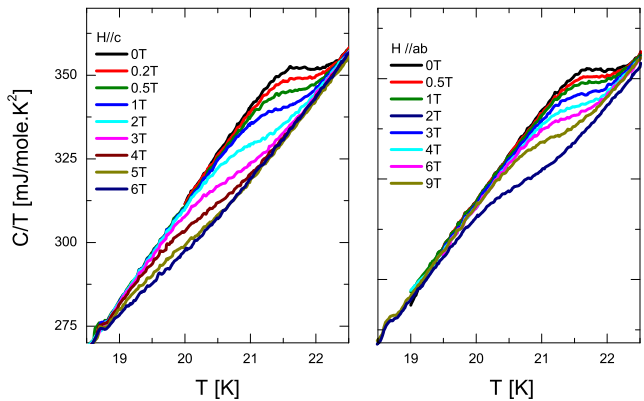}
	\caption{\label{jump} Temperature dependence of specific heat, represented on a $C/T$ vs $T$ plot, zooming into the range close to $T_c$ for magnetic fields $H \parallel c$ (left panel) and $H \bot c$ (right panel). 
}
	\label{jump}
\end{figure}

Fig.~\ref{jump} shows the temperature dependence of the specific heat obtained on a $m$=1.9~mg sample, used in magneto-optical and magnetization studies of the critical currents, Fig.~\ref{jump}. This sample is of very regular shape which allowed for its precise alignment in magnetic field. 

The jump in the specific heat at $T_c$ broadens with the increase of magnetic field for both directions of an applied magnetic field (Fig.~\ref{jump}). This makes the determination of $T_c(H)$ more ambiguous. To obtain the best resolved jump at the lowest fields, which are of main interest for our comparison, we subtract the $C/T$ data for $H(\parallel c)=6$~T from all low field curves, as shown in Fig.~\ref{CovT6Tsubtraction}. This subtraction removes the large non-superconducting background to the specific heat and reveals a sharper $C/T$ jump. To determine $T_c(H)$ we have used linear fits of the rising portion of the $C/T$ data, as shown in Fig.~\ref{CovT6Tsubtraction}. Thus we have determined the onset of the specific heat anomaly, while the maximum position was used as yet another criterion for the $T_c(H)$ determination. 
Depending on the criteria used, we get different slopes of the $H_{c2}(T)$ at $T_c$. For $H\parallel ab$ ($H \parallel c$) it ranges from 8.8~(3.4)~T/K from the onset of the $C/T$ jump to 5.3 (2.45)~T/K from the position of the maximum. 

\begin{figure}
	
		\includegraphics[width=8.5cm]{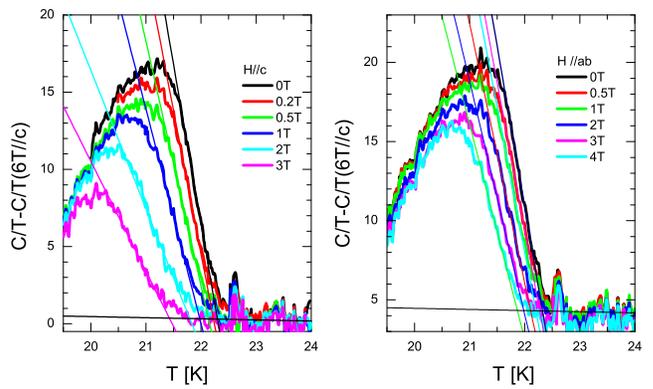}
	\caption{\label{CovT6Tsubtraction} Temperature dependence of the difference of specific heat, $C/T- C/T(H\parallel c,~6~T) $, for $H\parallel c$ (left panel) and $H\bot c$ (right panel). This subtraction removes non-superconducting background to the specific heat for $T>T_c(6T,~H\parallel c)$. The $T_c$ was determined from the position of the maximum in the data and from the position of an onset, determined as a crossing point of linear extrapolations of the rising part of the $C/T$ jump and of the constant data above $T_c$, as shown. 
}
	\label{CovT6Tsubtraction}
\end{figure}

\begin{figure}
	
		\includegraphics[width=8.5cm]{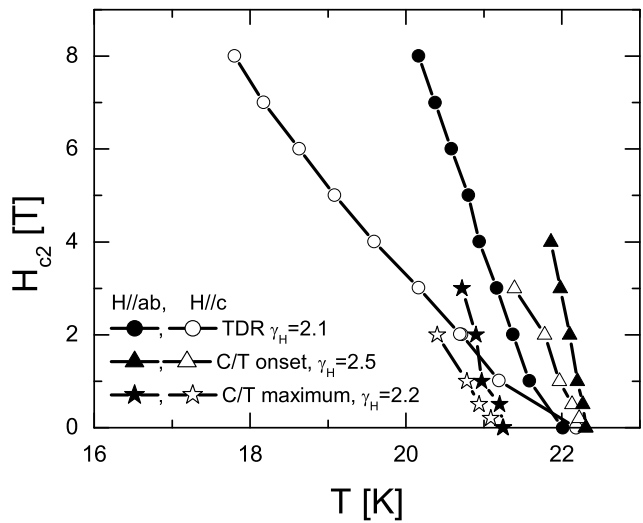}
	\caption{ Temperature dependence of the upper critical fields as determined from the onset (triangles) and the maximum position (stars) in specific heat data and from TDR data (circles). The data for $H \parallel ab$ is shown with solid symbols, for $H \parallel c$ is shown as open symbols. 
}
	\label{Hc2}
\end{figure}

\begin{figure}
	
		\includegraphics[width=8.5cm]{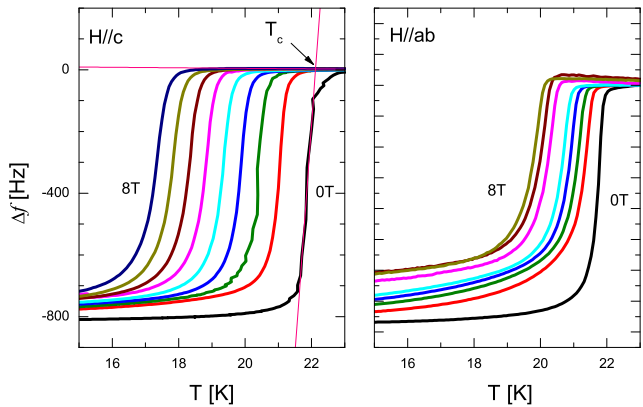}
	\caption{ Temperature dependence of the TDR frequency shift in magnetic fields applied along the tetragonal $c$-axis (left panel, $H\parallel c$) and perpendicular to it (right panel, $H\parallel a$). 
The superconducting transition temperature was defined as a crossing point of linear extrapolations of the shift in the normal state and at the slope.
}
	\label{TDRHc2}
\end{figure}

Fig.~\ref{Hc2} shows a comparison between bulk thermodynamic $H_{c2}$ determined by both criteria from the specific heat with $H_{c2}(T)$ determined from TDR measurements. The TDR data used for the $H_{c2}$ determination are shown in Fig.~\ref{TDRHc2}. 
In zero applied external magnetic field, the frequency shift is representative of the variation with temperature of the London penetration depth. In constant magnetic field it represents the onset of the superconducting shielding and is close to the onset of the resistive transition \cite{RTGordon}. The slopes determined from the TDR measurements are 4.56~T/K for $H \parallel ab$ and 1.87~T/K for $H \parallel c$. Thus by all the criteria used we have obtained $\gamma_H$ between 2.1 and 2.6, which matches within the error bars to the electrical resistivity anisotropy determined from the Montgomery technique.

\subsection{ Anisotropy of the London penetration depth}

\begin{figure}
	
		\includegraphics[width=8.5cm]{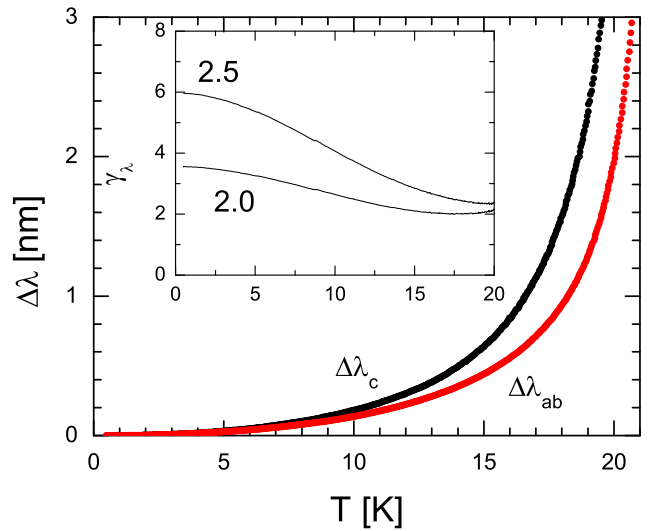}
	\caption{ Variation of the London penetration depth, $\delta \lambda \equiv \lambda(T) - \lambda(T=0)$ with temperature for two orientations of AC field with respect to the sample. Inset shows temperature dependence of the anisotropy of penetration depths, $\gamma _{\lambda}$, assuming $\gamma _{\lambda}= \gamma_H$ at $T=0.9T_c$ and two selected values of $\gamma_H$= 2.0 and 2.5. 
}
	\label{penetration_depth}
\end{figure}

The in-plane penetration depth, $\Delta\lambda_{ab}$(T), was determined by applying the rf magnetic field along the $c$-axis ( H$_{rf}||c$). In this geometry, the screening currents flow within the $ab$-plane, so only the in-plane component of the penetration depth is probed. When H$_{rf}\perp c$, the screening currents flow both in-plane and along the $c$-axis. Using the model described by Ref.~\cite{Prozorov2006}, the $c$-axis component $\Delta\lambda_{c}$ was obtained. Because the TDR technique measures changes in penetration depth with temperature ($\Delta\lambda$(T)), in order to calculate $\lambda$(T) we need as a reference, the absolute value of penetration depth at a certain temperature. For the in-plane component, we have used the zero temperature value, $\lambda_{ab}(0)\approx$ 208 nm, from Ref.~\cite{RTGordon}. As we are not aware of any previous reports of $\lambda_{c}$(0), for the $c$-axis component we make the assumption that near T$_c$, according to Ginzburg-Landau theory,  both the upper critical field and the penetration depth anisotropies should be equal. Considering $\gamma_{\xi}=\gamma_{\lambda}$, we set at T=0.9T$_c$, $\lambda_{c}=\gamma_H \times\lambda_{ab}$ with two selected values for $\gamma_H$= 2.0 and 2.5. The resulting low anisotropy and its weak temperature dependence are consistent with the overall picture of small anisotropy in the pnictides (Fig.~\ref{penetration_depth}).

\subsection{ Anisotropy of the critical current}

In Fig.~\ref{IV} we show isothermal current density- voltage, J-V, dependences for currents flowing along $a$-axis (left panel) and along tetragonal $c$-axis (right panel). The resistivity of the sample used in the critical current measurements along the $c$-axis at room temperature was about 800 $\mu \Omega cm$, a value consistent with minimal cracking. For both directions of the current flow we have determined the same resistive transition with the onset at 22.5~K, the midpoint at 22~K and the zero-resistance state at 21.6~K ($\pm$0.05~K in all cases). 
The VJ curves are linear in the normal state, implying that Joule heating is insignificant. The critical current was determined at the point of the sharpest voltage rise, for which the derivatives of the actual V-J curves were taken \cite{vortex}. This procedure is unambiguous when the sample is cooled somewhat below a temperature where its resistance becomes zero. In the inset of Fig.~\ref{critcurrents} we compare the critical current densities determined from the transport measurements. The data reveal only modest anisotropy (less than 1.5). However, the temperature range for the determination of the transport critical current density is limited to the immediate vicinity of $T_c$, and small differences in $T_c$ can potentially affect this ratio. To determine the critical current anisotropy in a broader temperature range we turn to the magnetic measurements. 

\begin{figure}
	
		\includegraphics[width=8.5cm]{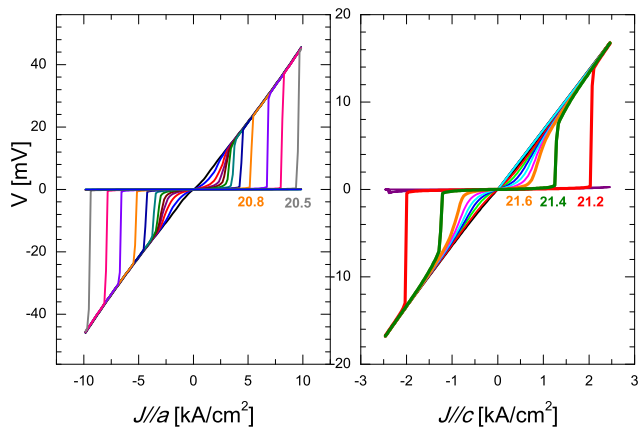}
	\caption{\label{IV} Isothermal dependences of voltage drop on the sample, $V$, vs current density, $J$, taken in the vicinity of the resistive transition to the superconducting state. Left panel, current along $a$-axis, right panel- current along tetragonal $c$-axis. For current densities higher than certain values the curves become linear. The critical current density was determined as the position of the maximum of the derivative of $V(J)$. 
}
	\label{IV}
\end{figure}

\begin{figure}
	
		\includegraphics[width=8.5cm]{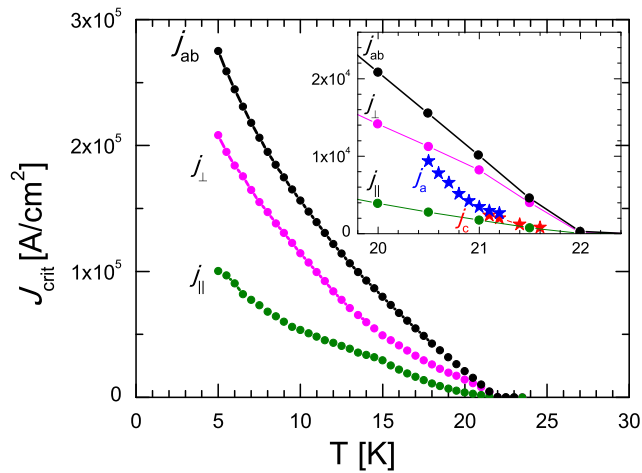}
	\caption{\label{critcurrents} Superconducting critical currents as determined from transport measurements (stars in the inset) and from magnetization measurements using the Bean model (main panel and inset, solid circles). 
}
	\label{critcurrents}
\end{figure}

We use the Bean model with field-independent, constant throughout the sample, supercurrent density $j_{s}$. Assuming that $j_{s}$ is isotropic in the crystallographic $ab-$ plane, three different current densities are possible depending on the orientation and the direction of motion of the Abrikosov vortices under the influence of the Lorentz force.

Let us consider a rectangular prism - shaped crystal of dimensions, $2c<2a<2b$. In the crystals of interest, the crystallographic $ab-$plane has the largest area and is parallel to the geometric $ab-$ plane of a prism. The smallest dimension, $2c$, is the sample thickness. When a magnetic field is oriented along the crystallographic $c-$ axis (along the $c-$ side), the measured magnetic moment is denoted by $M_{c}$. Similarly, the magnetic moments measured along the $b-$ side and the $a-$ side are $M_{b}$ and $M_{a}$, respectively. When a magnetic field is applied along the $c-$ side, the Abrikosov vortices are perpendicular to the $ab-$ plane and their gradients generate a supercurrent density, $j_{ab}$, given by%
\[
j_{ab}= \mathscr{M}_{c}\frac{20}{Va\left(  1-\frac{a}{3b}\right)  }=\frac{20}{a}%
\frac{M_{c}}{\left(  1-\frac{a}{3b}\right)  }%
\]
where $V$ is the sample volume and $M_{c}=\mathscr{M}_{c}/V$ is the volume magnetization. This formula is written in practical units where all lengths are in cm, magnetic moment is in emu and current density is in A/cm$^{2}$. 

The situation is more complicated when a magnetic field is applied parallel to the $ab-$ plane. Here we have two different current densities; one for the vortex motion across the planes, $j_{\perp}$, and another parallel to the planes, $j_{\parallel}$. By using the rectangular prism shaped samples with substantially different $a$ and $b$ sides, we can separate these two currents. Using the Bean construction, we have%
\[
M_{a}=\frac{j_{\perp}c}{20}\left(  1-\frac{c}{3b}\frac{j_{\perp}}%
{j_{\parallel}}\right)
\]
and,%
\[
M_{b}=\frac{j_{\perp}c}{20}\left(  1-\frac{c}{3a}\frac{j_{\perp}}%
{j_{\parallel}}\right)
\]


Solving for the currents we obtain,%
\[
j_{\perp}=\frac{20}{c}\frac{bM_{a}-aM_{b}}{b-a}%
\]
and%
\[
j_{\parallel}=\frac{20}{3ab}\frac{\left(  bM_{a}-aM_{b}\right)  ^{2}}{\left(
M_{a}-M_{b}\right)  \left(  b-a\right)  }%
\]
where both formulas are applicable as long as $aj_{\parallel}>cj_{\perp}$.
If this condition is violated, similar formulas can be easily obtained and the applicability of a particular expression can be checked by examining the
results.

Fig.~\ref{critcurrents} compares the critical currents obtained from transport and magnetic measurements. In a broad range of temperatures from $T_c$ down to 5~K, the magnetically determined in-plane and inter-plane critical currents are different by a factor of 3, slightly more than in the direct transport measurements. Fast magnetic relaxation in pnictide superconductors may play a substantial role and result in a noticeable difference between transport and magnetic currents \cite{vortex}. Therefore, the comparative agreement between the two estimates is quite good.

\section{Discussion}

\subsection{ Comparison with other iron pnictides}

We now turn to a comparison with other iron pnictide compounds. Until now, the anisotropy has been characterized reliably only for $\gamma_H$. Most directly this has been accomplished from a scaling entropy procedure in heat capacity measurements \cite{Welp1,Welp2}. For NdFeAsO$_{1-x}$F$_x$ this has given an anisotropy of $\gamma_H \sim$4, while for Ba$_{1-x}$K$_x$Fe$_2$As$_2$ it is $\gamma_H \sim$2.6. This is not very different from determinations from resistive measurements, as we discussed above. Anisotropy inside the superconducting state was determined from torque measurements at fields $H \ll H_{c2}$. This is sensitive to flux distribution in the sample (see magnetooptical imaging in \cite{vortex}), and studies performed until now give a scatter of values from about 9 in SmFeAs(O,F) \cite{Balicas} to 1.2 in PrFaAs(O,F) (see Ref. \cite{anisotropy_japanese} for a summary of all data). By looking at different properties, our study reveals that the small anisotropy is typical for all normal and superconducting properties of the iron pnictides. 

The fact that the anisotropies found in this study for the normal state resistivity and for the slopes of the upper critical fields match, implies an orbital mechanism behind superconductivity suppression with magnetic field. Provided the same is the case in other superconductors in this family, high values of the $H_{c2}$ then would imply that the Fermi velocities are small, and three-dimensional portions of the Fermi surface play an important role in the band structure of these materials. 

It is interesting to understand whether or not a small anisotropy favors a higher $T_c$. A long history of superconductivity research in low dimensional materials following the original idea by Little \cite{Little}, suggests that a two-dimensional electronic structure is favorable for the realization of a higher $T_c$, see for example Ref.~\cite{Ishiguro}. For magnetically mediated superconductivity a direct link between the anisotropy and the $T_c$ is suggested \cite{Mathur}. We now turn to compare the anisotropy of the upper critical fields in different iron arsenide materials. In the Co-doped version of BaFe$_2$As$_2$ with $T_c$=22.5~K the anisotropy is about 2 from resistive \cite{Athena,NiNiCo} and 2.1 to 2.6 from specific heat (Fig.~\ref{Hc2}) determinations. In Ba$_{1-x}$K$_x$Fe$_2$As$_2$ with $T_c$=30K $\gamma_H$ is in the range of 2 to 3.5 from the resistive \cite{NiNi} and equals 2.6 from specific heat measurements \cite{Welp2}. In the 1111 compound NdFeAs(O,F) with $T_c \approx$50~K it is about 4 from specific heat \cite{Welp1} and TDR \cite{Martin} measurements. We may notice that the anisotropy is increasing in parallel with $T_c$. This may be an interesting trend that is worthy of further exploration. It actually reveals the underlying trend in the evolution of the band structure of these compounds. 

\subsection{ Band structure}

A key structural feature of the iron arsenides, which distinguishes them from the cuprates and makes a profound effect on their electronic structure, is the location of the As atoms above and below the layer of Fe atoms. We recall that in the cuprates oxygen atoms are located in the planes. 

\begin{figure}
\includegraphics[width=8.5cm]{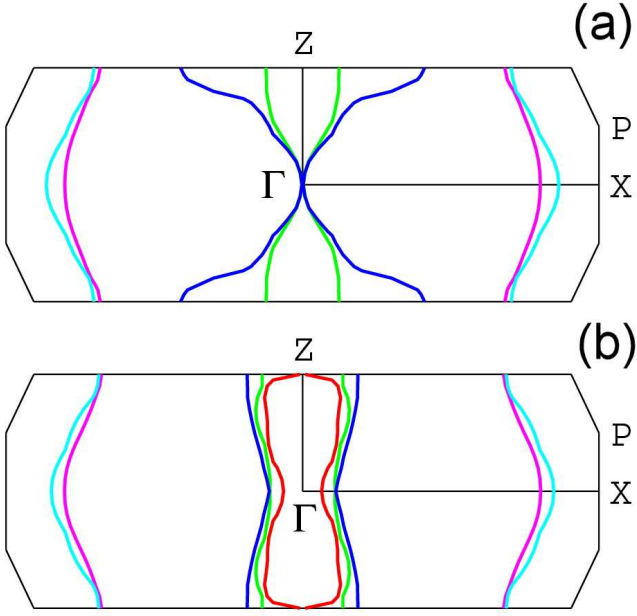}
\caption{ Fermi surfaces of Ba(Fe$_{1-x}$Co$_x$)$_2$As$_2$, $x$=0.074,  calculated assuming relaxed (a, top panel) and experimental (b, lower panel) positions of As atoms in the lattice. 
}
\label{Fermi_surface}
\end{figure}

Since actual location of As atoms can vary, in our calculations we used two positions. In the first one, the position of As atoms was determined from the minima of total energy. We refer to this as relaxed position below. The calculated position of As atom, $z_{As}$=0.341, is very close to that found in previous calculations, $z_{As}$=0.342 \cite{MP2}, but is significantly lower than the experimental value of $z_{As}$=0.355 \cite{latticeconstants}. To simulate 0.074 Co doping we replaced Fe atoms by virtual atoms having $Z$=26.07 (virtual crystal approximation). The Fermi velocities, calculated using the Bolz-Trap \cite{BolzTrap} package, are very sensitive to the As positions. A similar trend is found in magnetic properties \cite{MP2,MP1}. The downshift by 0.16 \AA\ increases both the band dispersion and the Fermi velocity along the $z$ direction, see Fig.~\ref{Fermi_surface}. The calculations with relaxed $z_{As}$  give $V_{Fa}^2/V_{Fc}^2$=3 for pure BaFe$_2$As$_2$ and 2.7 for the $x$=0.074 doped compound studied here. 
Using the experimental $z_{As}$ we come to a much larger anisotropy of 12.1 and 9.0, for pure and doped materials, respectively.

This difference in anisotropy is affecting the most topology of the sheets of the Fermi surface surrounding $\Gamma$ point in the Brillouin zone (Fig.~\ref{Fermi_surface}). Cylinders around $X$ point remain warped and do not change much with variation of As atom positions.

As we see, both positions of As give anisotropies not very different from those that have been experimentally determined. More precise determination of the As position in doped materials as well as anisotropy study of the parent pure compound are necessary to clarify the situation.

\section{Conclusions}

Iron arsenic superconductors reveal small anisotropies of the electronic structure and, as a result of this, of the superconducting state. Since many theories consider high anisotropy as an important ingredient for the achievement of high transition temperatures, this experimental observation puts strong constraints on the models suitable for the explanation of superconductivity in these exciting compounds.


We thank A. Kaminski and Y. Lee for discussions and M. Kano for inspiration. M.A.T. acknowledges continuing cross-appointment with the Institute of Surface Chemistry, National Ukrainian Academy of Sciences. Work at the Ames Laboratory was supported by the Department of Energy-Basic Energy Sciences under Contract No. DE-AC02-07CH11358. R. P. acknowledges support from Alfred P. Sloan Foundation.



\begin{references}
\bibitem{linkedgammas}
The anisotropic conductivity in the normal state is governed by the tensor $\langle v_iv_j \tau\rangle$ where $v_i$ is the Fermi velocity and $\tau $ is the scattering time.  Assuming that $\tau$ is nearly isotropic, one obtains $\gamma_\rho= \langle v_c^2 \rangle/\langle v_a^2 \rangle$. On the other hand, superconducting anisotropy near $T_c$ is governed by the inverse ``mass tensor" calculated in Ref.\cite{Gorkov} for clean superconductors:   $m_{ij}^{-1} \propto \langle\Delta^2 v_iv_j  \rangle$. This tensor   determines   anisotropies of the upper critical field and of the penetration depth:
\begin{equation}
\gamma_H^2 = \gamma_\lambda^2 = \langle\Delta^2 v_a^2  \rangle/\langle\Delta^2 v_c^2  \rangle  \,.
\label{gH=glambda}
\end{equation}
Strong scattering erases the effect of whatever anisotropy of $\Delta$ \cite{Pokrovsky,KRap} upon $ \gamma_\lambda$ and $ \gamma_H$ and we have in the dirty limit:
\begin{equation}
\gamma_H^2 = \gamma_\lambda^2 =  \langle  v_a^2  \rangle/\langle  v_c^2  \rangle  \,.
\label{dirty}
\end{equation}
\end{references}
\end{document}